# A Model for Evaluating Algorithmic Systems Accountability

Yiannis Kanellopoulos,
Code4Thought
P.O. Box 26441
Greece
yiannis@code4thought.eu

## ABSTRACT

Algorithmic systems make decisions that have a great impact in our lives. As our dependency on them is growing so does the need for transparency and holding them accountable. This paper presents a model for evaluating how transparent these systems are by focusing on their algorithmic part as well as the maturity of the organizations that utilize them. We applied this model on a classification algorithm created and utilized by a large financial institution. The results of our analysis indicated that the organization was only partially in control of their algorithm and they lacked the necessary benchmark to interpret the deducted results and assess the validity of its inferencing.

## 1 PROBLEM DESCRIPTION

There is little doubt that algorithmic systems are making decisions that have a great impact on our daily lives. Or as [1] notes, *authority is increasingly expressed algorithmically* and decisions that used to be based on human intuition and reflection are now automated [2]. So, transparency over how these systems work matters not as an end in itself but merely as means towards accountability.

By accountability we mean the degree of which one decides when and how an algorithmic system should be guided (or restrained) [2] in the risk of crucial or expensive errors, discrimination, unfair denials, or censorship.

According to [11] algorithmic systems are merely defined as a combination of data structures and operations on them. However, in this paper we use this term (algorithmic systems) in order to express the influence of the organization utilizing (or creating) an algorithm and for that they need to cater for accountability on the algorithm itself that has to be designed for that purpose. Our thesis is that the mandate for transparency and for accountability should be imposed to both of them. An organization that considers accountability and an algorithm designed as such can gain and have respectively the following benefits:

● There can be trust between the organization utilizing the algorithms and those affected by its output (be it clients, citizens or simple users), since the results can be explained,
Improving the algorithm's output, since identified weighting factors and thresholds can be calibrated/fine-tuned if needed,
● Rendering the algorithms more persuasive, since their reasoning will be easier to explain.

This is not an easy task and the main challenges are the following:

● In several cases precision is preferred compared to transparency. For instance, in the case of deep learning we need to typically comprehend the relationships among thousands of variables computed by multiple runs through vast neural networks. A task that is impossible for human brain.
● Organizations tend to keep their algorithms secret claiming they want to preserve valuable intellectual property or avoiding the risk of them getting gamed (e.g. in the case of credit scoring algorithms),
● There is no widely accepted industry-standard that defines how an algorithmic system should be evaluated in terms of transparency and accountability and what are the suggested criteria.

The primary objective of this paper is to present an evaluation model whose purpose is to assess the transparency and accountability of an algorithmic system (i.e. algorithm's design as well as an organization's maturity). It is based on the work of [3] that mainly focuses on the news and media domain, and [5] that is geared towards how organizations provide for accountability.

The main characteristics of this model are:

- It is business-domain and technology agnostic, so it can be operationalized at any type of organization or business and algorithm,
- It is not intrusive as it doesn't require any data or input risking to disclose the specifics of an algorithmic system. It merely consists of a set of questions that require experts' input.

We did apply this model in a large financial institution that wanted to develop a decision support system for their web and mobile banking platforms.

In Section 2 we present our model, while in Section 3 we discuss the results and the conclusions from our case study. Finally, at Section 4 we discuss the next steps of our research.

## 2  ACCOUNTABILITY MODEL DESCRIPTION

In the figure below, we present our model whose scope includes the algorithms themselves as well as the organization which utilizes them and needs to cater for their accountability.

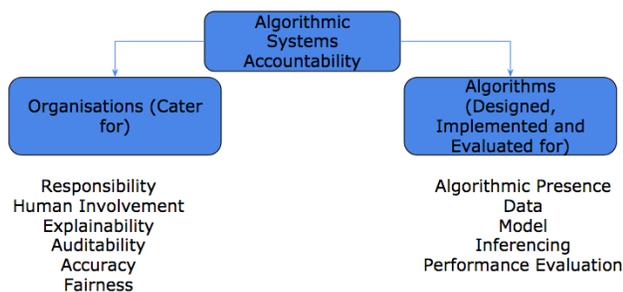

**Figure 1: Organizational and Algorithmic Accountability**

## 2.1 Technical Perspective: Algorithms

### 2.1.1 Algorithmic Presence:

This characteristic concerns the disclosure if and when an algorithm is being employed at all. The goal for this characteristic is to identify what is the problem that needs to be solved and given its nature, the type of the employed algorithm.

### 2.1.2 Data:

Data is probably the most crucial part of an algorithmic system as it is usually the place where bias can be introduced. They are being characterized mainly by the following aspects:
- Quality, which involves their accuracy, completeness, and uncertainty, as well as their timeliness, representativeness of a sample for a specific population, and assumptions or other limitations.
- Handling, which includes data definitions, ways of collection, vetting and editing (manually or automated).

Further indicative questions we have used for our model regarding this aspect were:
- How are various data collected and annotated?
- How are various data labels gathered?
- How are missing values being handled?
- Which elements are being filtered away and why?
- How do they reflect a subjective or objective process?
- Do incorporated dimensions have personal implications if disclosed?

### 2.1.3 Model:

It involves the model or algorithm itself as well as the process followed for its construction. The goal here is to identify its input, the selected features or variables along with their weights (in case they are weighted).

### 2.1.4 Performance Evaluation:

This concerns the creation of a framework for choosing the right algorithm and for evaluating how well it performed. It mainly consists of the selection of the appropriate metrics for this purpose. These influence how one weights the importance of different characteristics in the results and their ultimate choice of which algorithm to choose. So, what we're interested in identifying here are the employed evaluation metrics, the reasoning behind their selection and most importantly how these are being utilized and interpreted.

### 2.1.5 Inferencing:

It is related to the algorithm's evaluation in terms of its accuracy and error margin and its creator's ability to benchmark it against standard datasets and standard measures of accuracy. Points we take into consideration here are the error margin, the accuracy rate, the false positives versus the false negatives, the remediation process when errors occur and finally the ability to identify the root cause of an error (i.e. human involvement, data inputs, or the algorithm itself?).

## 2.2 Organisational Perspective: Accountability Guiding Principles

### 2.2.1 Responsibility:

Organizations need to have publicly (or at least externally) available visible ways of redress for adverse individual or societal effects. What is more, they need to designate an internal role for the person who is responsible for the timely remedy of such issues.

### 2.2.2 Human involvement:

It requires explaining the goal, purpose, and intent of the algorithm. Here, we're interested in knowing the person responsible if users are harmed by the algorithm, the reporting and correcting process, who will be having the authority to decide on necessary changes and finally what are the roles of people who have direct control over the algorithm.

*2.2.3 Explainability:*

Algorithmic decisions as well as any data driving those decisions should be explained to end-users and other stakeholders in non-technical terms. It is important here to identify who are the end-users and stakeholders of the employed algorithm, how much of it can be explained to them and how much of the data sources can be disclosed.

*2.2.4 Accuracy:*

Sources of error and uncertainty throughout the algorithm and its data sources should be identified, logged and articulated so that expected and worst-case implications can be understood and trigger mitigation procedures. It is important to know the sources of error and how their effect is mitigated, the level of confidence on decisions output, the veracity of the data sources of the algorithm and finally if there are in place any risk management processes in case of a worst-case scenario.

*2.2.5 Auditability:*

An accountable organization should enable interested third parties to probe, understand, and review the behavior of their algorithms through disclosure of information that enables monitoring, checking, or criticism, including through provision of detailed documentation, technically suitable APIs, and permissive terms of use. It is important to know if they can provide for public auditing (i.e. probing, understanding, reviewing of system behavior) or if there is sensitive information that would necessitate auditing by a designated 3rd party. In case of the latter, organizations need to define how they plan to facilitate such public or third-party audit without any opening of the algorithm to unwarranted manipulation.

*2.2.6: Fairness:*

An organization that caters for accountability should ensure that algorithmic decisions do not create discriminatory or unjust impacts when comparing across different demographics (e.g. race, sex, etc.). So, it is important to define whether there are any particular groups which may be advantaged or disadvantaged, in the context which they are deploying, their algorithms. What is more, organizations need to define and quantify what is the potential damaging effect of uncertainty / errors to different groups.

## 3   RESULTS

We applied our model in order to evaluate a classification [7] algorithm of a large financial institution. The goal of the algorithm is to classify the users of a mobile and web platform based on how digitally literate they are. The organization is trying to improve its users' experience by providing advanced functionality to the literate ones and basic to those who are classified as non-literate.

By applying the model, our analysis indicated that the organization was partially in control of its algorithm. When it comes to the implementation of the algorithm itself, the main issue was the inability of reliable inferencing. The reason for this was the lack of a benchmark that the organization could employ to interpret the deducted results and assess their validity. The main findings from our analysis were:

● The organization has full control over the quality and the selection of the data used for feeding the algorithm. On the other hand, there seems to be no formal processes for handling any issues caused by the algorithm (e.g. harmed or disappointed/frustrated users)

● There is partial control over the algorithm design as the team decided to employ a set of classical data mining techniques such as clustering [8] and association rules [9], [10] that are easier to explain compared to a neural network.

● However, for the task at hand, that is to classify users based on their digital literacy, the selected approach wasn't solving it adequately as the created algorithm cannot provide reliable inferencing. The main reason for this is the lack of labeled data and of the ability to interpret the derived clusters and rules using human judgement. Nonetheless, state-of-the-art semi-supervised techniques like [12,13] could be employed if an internal domain expert had annotated a few initial training examples.

## 4   CONCLUSIONS AND FUTURE WORK

In summary, we have performed both a theoretical and experimental study on how we can develop a model that evaluates how accountable an algorithmic system is (i.e. algorithm and the organization that utilizes it). The application of the model in a real-world situation and more specifically in a financial institution proved to be of value for the following reasons:

● Questions were deemed practical and helpful since they were challenging the algorithm's creators for their choices,

● It helped identifying areas for improvement at the system as well as at the algorithmic level.

● It provided guidance on what are the areas of attention when designing the algorithm.

● Similarly, that was the case for the organization by defining the lack of a role/person of responsibility for making sure the algorithm works as planned.

As next steps for the model's development, we identified the following:

● The current model is evaluating accountability in a qualitative way. Especially for auditing and governance purposes it might be useful to further research the possibility to quantify and measure the level of accountability for both the algorithm and the organization in question,

- Apply the model in more industrial cases in order to create a benchmark that can be used for the purposes of evaluating the accountability of algorithmic systems,
- Utilize a tool that will facilitate the explainability of a given algorithmic system.